\documentstyle[11pt,epsfig]{article}

\def\be{\begin{eqnarray}}
\def\ee{\end{eqnarray}}

\def\roughly#1{\mathrel{\raise.3ex\hbox{$#1$\kern-.75em%
\lower1ex\hbox{$\sim$}}}}

\def\gsim{\roughly>}
\def\la{\langle}\def\ra{\rangle}

\def\del{\partial}
\def\Tr{\rm Tr}
\def\tr{\rm tr}
\begin{document}

\begin{titlepage}

\hfill {\today }

\begin{center}

\centerline{\Large\bf Dilatons in Hidden Local Symmetry}

\centerline{\Large\bf for Hadrons in Dense Matter}

\vspace{.30cm}

Hyun Kyu Lee$^{a}$ and Mannque Rho$^{a,b}$

\vskip 0.50cm

{(a) \it Department of Physics, Hanyang University, Seoul 133-791, Korea}

{(b) \it Institut de Physique Th\'eorique, CEA Saclay, 91191 Gif-sur-Yvette, France}
%{\it 91191 Gif-sur-Yvette, France}

\end{center}
\vskip 1cm

\centerline{\bf Abstract}
\vskip 0.5cm

With the explicit breaking of scale invariance by the trace anomaly of QCD rephrased in
terms of spontaneous breaking, low-energy strong interaction dynamics of dense (and also
hot) matter can be effectively captured by -- in addition to the Nambu-Goldstone bosons
and the vector mesons -- two dilaton fields, the ``soft" ($\chi_s$) field that is locked
to chiral symmetry and the ``hard" ($\chi_h$) field which remains unaffected by chiral symmetry.
The interplay of the soft and hard dilatons plays a subtle role in how chiral symmetry is manifested
in hot and/or dense matter. The scale anomaly in which the soft component intervenes vanishes at
the chiral transition in a way analogous to the restoration of scale symmetry in the Freund-Nambu
model, while that of the hard component remains broken throughout the QCD sector. Most remarkable
of all is its role in the chiral anomaly sector through a ``homogeneous Wess-Zumino (hWZ) term" of
the form $\omega_\mu B^\mu$ on the structure of a single baryon as well as dense baryonic matter. It figures crucially in predicting a ``Little Bag" for the nucleon and a ``quarkyonic phase" in the form of a half-skyrmion matter at high density. We show how the vanishing of the vector-meson mass at the vector manifestation fixed point in hidden
local symmetry theory can be related to the property of the ``matter field" in the Freund-Nambu model
that leaves scale symmetry invariant. The emerging structure of dense hadronic matter in the model so constructed  suggests what could be amiss in describing dense matter in holographic dual QCD at its large $N_c$ and 't Hooft limit.

\end{titlepage}\vskip 0.5cm

\section{Introduction}
The trace anomaly in QCD is associated with the emergence of a scale at the quantum level in
a theory that has no scale at the classical level. QCD also has it that chiral symmetry is broken
giving rise to the hadron mass. Thus the chiral symmetry breaking ($\chi$SB) and the scale symmetry
breaking (SSB) are intricately tied to each other. If one ignores the quark mass, then scale
invariance is broken explicitly by the trace anomaly while chiral symmetry is broken spontaneously
at low energy. How are these related? This is an issue which remains unclarified in the effort to
unravel the mechanism of mass generation for light-quark hadrons such as the nucleon, the vector
 mesons etc.

It has recently been observed that the dilaton (or rather dilatons) associated with breaking scale
invariance through the trace anomaly of QCD play a dramatic role for the nucleon structure, the phase
structure of dense matter~\cite{PRV-dilaton}, and the properties of hadrons in dense medium, such as
the pion decay constant, the vector meson mass and the repulsive force in nuclear interactions~\cite{LRS}.
 We would like to understand the apparent dichotomy in the description of the scaling behavior of
 hadrons in medium, namely that the scaling property of hadrons in medium as a function of temperature
 or density can be attributed to the behavior of the dilaton condensate~\cite{BR91} on the one hand and
 as the renormalization group flow of hidden local symmetry toward the vector manifestation fixed
 point~\cite{HY:PR} on the other hand. While the former involves scalar fields, the latter involves (
 hidden) gauge fields. The question is: How are these related in nuclear medium?

In this paper, we address the connection between the chiral symmetry and the scale symmetry expressed
in terms of the quark condensate for the former and the gluon condensate for the latter as well as
their ramification on chiral symmetry restoration and possible meson condensation such as kaon
condensation in dense medium.

Given a Lagrangian, a symmetry can be explicitly broken by either a breaking term put in the (effective)
Lagrangian or through quantum anomaly of the theory with the Lagrangian that is scale-invariant. QCD
has explicit symmetry breaking by the latter, namely the trace anomaly. In our discussion, we will follow
the first procedure by transforming the explicit breaking into a spontaneous breaking and put in the
Lagrangian. Thus we will be dealing with the situation in which the scale symmetry is broken both
explicitly and spontaneously. In this way we will be able to connect $\chi$SB which generates masses to
scale symmetry breaking which also generates masses. This would
allow us to connect the spontaneously broken chiral symmetry to the dilaton associated
with spontaneously broken scale symmetry and to other condensates such as kaon condensate.

By a standard procedure, one can phrase the trace anomaly of QCD -- which represents an {\em explicit
breaking} by quantum effects of the scale symmetry -- in terms of a {\em spontaneous breaking}~\cite{schechter,migdal-shifman}.
We introduce two dilaton fields $\chi_s$ and $\chi_h$ representing respectively ``soft" and
``hard" components~\cite{miransky} ignoring for the moment the possible mixing between the
two~\footnote{The possibility of an intricate mixing between the two in the vicinity of chiral
transition will be commented on below.}. Translating the explicit breaking of scale invariance
into a spontaneous breaking~\footnote{This procedure could be justified by introducing a dynamical
field for the QCD scale parameter $\mu$ that figures in the dimensional regularization and by
calculating the Coleman-Weinberg-type potential involving that field of the resulting theory.
See ~\cite{shapo}.}, one can write the trace of the energy-momentum tensor in terms of the
condensate of the dilaton. The scale invariance broken spontaneously in that way can then be
restored with the dilaton mass going to zero at some scale in a manner closely resembling the
Freund-Nambu model~\cite{nambu,Zumino}. In our approach guided by lattice gauge calculations as
the chiral restoration point is approached, we are led to the scenario that the ``soft" dilaton
mass goes to zero, thereby restoring the ``soft" scale symmetry \`a la Freund-Nambu, whereas the
``hard" dilaton will stay massive across the chiral transition, the ``hard" scale symmetry
presumably getting restored at some higher (grand-unification?) scale. In \cite{shapo}, this
scenario is invoked for the hierarchy problem as well as cosmology all the way to the Planck
scale. In our paper, we will only consider the two lowest scales involved in QCD.

This paper answers two questions that remained unanswered up to date and pinpoints to a source of difficulty in holographic dual (gravity) approach to dense hadronic matter that is encountered in the large $N_c$ and 't Hooft limit.
One of the unanswered questions is the justification
of the scaling behavior for the homogeneous Wess-Zumino term (hWZ for short) of the
$\omega_\mu B^\mu$ type that played a crucial role in \cite{PRV-dilaton} for both the nucleon
structure and the property of dense matter in the skyrmion description. It is shown here
how the scaling assumed for the hWZ term arises in a scale-invariant form involving the two dilaton fields. The other answer is the connection between the restoration of the spontaneously broken ``soft" scale symmetry given
in terms of the $\la\chi_s\ra$ going to zero -- which figures in BR scaling~\cite{BR91} -- and
the vector manifestation fixed point where the gauge coupling constant $g$ goes to zero in hidden
local symmetry theory~\cite{HY:PR}. The first clarifies the possible conundrum in applying the Sakai-Sugimoto holographic dual model~\cite{sakai} -- which has a correct chiral symmetry structure -- to properties of hadrons in dense matter~\cite{kimetal,bergmanetal,rozalietal}: Certain meson properties behave contrary to what one expects in QCD as density is increased. The second provides a background for the notion of ``hadronic freedom" which has been suggested as a possible understanding of dilpeton production in hot or cold environment~\cite{BHHRS}.

\section{Scale Symmetry Breaking and Condensation}
To set up the web of intricate connections we wish to explore, we start with a general discussion on the breaking of scale symmetry and its order parameter~\cite{coleman}. Here our discussion will be nothing more than standard but it will serve purpose for defining the quantities that we will be concerned with.
\subsection{Scalar field theory}
We first consider the scalar field theory with the Lagrangian given by
\be
 {\cal L} = \frac{1}{2}\partial^{\nu} \phi\partial_{\nu} \phi -
V(\phi).
\ee
The divergence of a dilatation current is the trace of the (improved) energy momentum tensor of the
theory:
   \be
   \partial^{\mu}D_{\mu} =  \Theta^{\nu}_{\nu},
   \ee
where
   \be
   \Theta^{\mu \nu} = T^{\mu \nu} +
   \frac{1}{6}(\partial^{\mu}\partial^{\nu}-g^{\mu
   \nu}\partial^{\rho}\partial_{\rho})\phi^2, \label{mtensor}
   \ee
and $T^{\mu \nu} = -g^{\mu
   \nu} {\cal L} + \frac{\delta {\cal L}}{\delta \partial_{\mu} \phi}\partial^{\nu} \phi   $ is the
   conventional energy-momentum tensor. Now using the equation of motion for the scalar field,
we get
   \be
\Theta^{\nu}_{\nu} = 4V(\phi) - \phi \frac{\partial V}{\partial
\phi}.
   \ee
If $\Theta^{\nu}_{\nu}$ is non-vanishing, the scale symmetry is broken.  If we decompose
the potential into two parts, a symmetric and a symmetry breaking
part,  $V = V_S + V_{SB}$,
   \be
\Theta^{\nu}_{\nu} &=& 4V(\phi) - \phi \frac{\partial V}{\partial
\phi} ,\\
&=& 4V_{SB}(\phi) - \phi \frac{\partial
V_{SB}}{\partial \phi},\\
 \ee
where the second equality follows due to the vanishing of the symmetric part:
 \be
(\Theta^{\nu}_{\nu})_S= 4V_{S}(\phi) - \phi \frac{\partial
V_{S}}{\partial \phi} =0. \label{traces}
 \ee
The space-time-independent classical solution, $\phi_0$, is obtained by
 \be
\frac{\delta V}{\delta \phi}|_{\phi_0} =0. \label{phi0}
 \ee
If we take it as a choice of the vacuum with an expectation value of
$\phi_0 \neq 0$,  it is called ``$\phi$ condensation."

In the case where there is no $V_{SB}$, i.e., $V = V_S$,   $\phi_0$ is determined by \be \frac{\delta
V_S}{\delta \phi}|_{\phi_0} =0. \label{phi0s}\ee
Thus
 \be
\la(\Theta^{\nu}_{\nu})_S\ra = 4V_{S}(\phi_0),\label{traces1}
 \ee
Hence
$V_{S}(\phi_0)$ should vanish according to Eq.(\ref{traces}). This
indicates that, when we use the modified energy momentum tensor,
Eq.(\ref{mtensor}), the condition for the scale symmetric
potential, $V_S$,  is \be V_{S}(\phi_0)=0, \ee with
Eq.(\ref{phi0s}).  This is of course the statement that a non-zero vacuum energy breaks scale
invariance. One well-known example is $V_S = g\phi^4$, where $\phi_0 =0$ and $V_S(\phi_0)=0$.

Now if a nontrivial condensate, $\phi_0 \neq 0$, is possible only in the presence of
$V_{SB}$, this implies that the explicit symmetry breaking of scale symmetry is
responsible for the $\phi$ condensation which can be associated with spontaneous breaking of
the symmetry. The vacuum expectation value of the trace of the energy-momentum tensor is then given by
\be \la\Theta^{\nu}_{\nu}\ra = 4V(\phi_0)=4[ V_S(\phi_0) +
V_{SB}(\phi_0)], \label{vtrace}\ee where we use Eq.~(\ref{phi0}).
For a model with $4V(\phi_0)=0$,  the scale symmetry appears to be ``restored"
at the condensation. This is what happens in  the Freund-Nambu model reviewed below.
\footnote{If $V_{SB}$ is non-zero so that scale symmetry is explicitly broken in the Lagrangian,
this seems to mean that the explicit symmetry is restored by the vacuum! This seems like the
situation similar to what we are thinking
when we say kaons condense to ``restore broken flavor symmetry" to which we will return in a
later publication.}
\subsection{The Freund-Nambu model}
\subsubsection{Dilaton field $\phi$ and a matter field $\psi$}
Our principal thesis relies on the intricate structure of the class of Freund-Nambu-type models,
so let's start with a short review of the model.

The Freund-Nambu model has two real scalar fields, $\psi$ and $\phi$, with the potential
 \be
V(\psi, \phi) = V_a + V_b,\ee
where \be V_a &=& \frac{1}{2} f^2 \psi^2 \phi^2 \label{fns} \\
V_b &=& \frac{\tau}{4}[\frac{\phi^2}{g^2} - \frac{1}{2} \phi^4
  - \frac{1}{2g^4}]= \frac{\tau}{8g^4}(g^2\phi^2 -1)^2 \label{fnsb}.
  \ee
$V_a$ is scale invariant and $V_b$ consists of a scale invariant
term $ ~ \phi^4$ and two symmetry breaking terms $\frac{\phi^2}{g^2}$ and $- \frac{1}{2g^4}$.
Introducing the new field $\chi$,
 \be
\chi = (g^2\phi^2 -1)/(2g)
 \ee
with the scale transformation
\be
\delta \chi = \epsilon(x \cdot \partial +2) \chi +
\frac{\epsilon}{g},
 \ee
the potential can be written as
 \be
V(\psi, \phi) = \frac{1}{2} m^2_{\psi}\psi^2(1+2g\chi) +
\frac{1}{2} m_{\chi}^2\chi^2, \ee
where
 \be m^2_{\psi} =
\frac{f^2}{g^2}, ~~~~ m^2_{\chi} = \frac{\tau}{g^2}.
 \ee
In terms of the condensate $\phi_0$ given by the minimum of the potential, $\phi_0=1/g$,
the masses take the form
 \be m^2_{\psi} =f^2\phi_0^2, ~~~~ m^2_{\chi} = \tau\phi_0^2.
 \ee
It should be noted that the mass of $\psi$ comes from the scale invariant
term $V_a$  and is independent of $\tau$ so that it can
have an arbitrary value.  However $m_{\chi}$ depends on $\tau$, going to zero linearly as
$\tau\rightarrow 0$.
This is known to be a characteristic of an approximate
spontaneously-broken scale symmetry~\cite{Zumino}:  Only the mass of one scalar
field (Goldstone boson) must be small but the masses of all other fields can be arbitrary.

As announced above and will be seen below, the vacuum expectation
value of $\Theta^{\nu}_{\nu}$ (which measures the scale symmetry
breaking) turns out to vanish at the condensation, $\phi_0\neq 0$
(with the potential (\ref{fnsb}), $\phi_0=1/g$). As we know from
lattice calculations (recalled below), this is not consistent with
QCD since it is known to be non-vanishing with a non-vanishing
condensation, at least in temperature up to a few times the chiral
transition temperature $T_c$. One can, however, add a constant to
get $\Theta^{\nu}_{\nu} \neq 0$ for  $\phi_0$ without changing the
structure of spontaneous symmetry breaking of the scale symmetry.

Suppose it models a physical situation where the condensate $\phi_0$
which is $\frac{1}{g}$ at the classical level is driven to zero by
density or temperature. Then the masses of $\chi$ and $\psi$ scale
$\rightarrow 0$ for fixed $f$ and $\tau$. Thus the ``matter field"
$\psi$ gets driven massless as the dilaton $\chi$ becomes
massless. This is what will happen with the case discussed in
Section \ref{CS}. For instance, the $\rho$-meson mass going to
zero in hidden local symmetry in the vector manifestation where
the gauge coupling $g$ figures is analogous to the behavior of the
$\psi$ field where the condensate going to zero drives its mass to
zero.

\subsubsection{The dilaton ($\phi$) sector}

The potential for the dilaton $\phi$ field in the Freund-Nambu model (\ref{fnsb}) consists of three parts
with scale dimension $d$=2, 4 and 0 ($V_2$, $V_4$ and   $V_0$ respectively in $V_d$ convention),
  \be
  V_b(\phi) = V_2 + V_4 + V_0
  \ee
with
 \be
 V_2 &=& \frac{\tau \phi^2}{4g^2}, ~~ V_4=-\frac{\tau}{8}
 \phi^4, ~~
  V_0=-\frac{\tau}{8g^4}.
  \ee
The potential $V_4$ is scale-invariant and others break scale symmetry. In the notation used above,
  \be
  V_S = V_4, ~~  V_{SB} = V_2 + V_0.
  \ee
Let us consider a potential in the form of polynomial of   $\phi$
  \be
  V_b = \sum_d V_d,
  \ee
where only $V_4$ is scale invariant.
The divergence of the dilatation current gives
  \be
  \Theta^{\nu}_{\nu}= \sum_d (4-d) V_d \label{trace}
  = (4-2)V_2 + (4-0)V_0.
  \ee
The RHS is given by the explicit symmetry breaking terms.

The minimum of the potential is given by the condition
  \be
\sum_d d ~ V_d|_{\phi_0} =0. \label{vphi}
 \ee
If there is no explicit symmetry breaking,  $\phi_0$ cannot develop a non-zero VEV and hence there is
no spontaneous breaking of scale symmetry.  For $\phi_0$ to be condensed, $V_{SB}$ is needed.
For the Nambu-Freund model, $[2 V_2
+ 4 V_4]|_{\phi_0} = 0$, so the condensate is $\phi_0 = \frac{1}{g}$.
Using Eq.~(\ref{vphi}),  Eq.~(\ref{trace}) at the condensation can be written
  \be \sum_d (4-d) V_d|_{\phi_0} = -[\sum_d d ~ V_d|_{\phi_0}] + 4 V(\phi_0) =
  +4V(\phi_0),
  \ee
as in Eq.~(\ref{vtrace}).  One can choose a constant term, $V_0$, in the potential to make the
expectation value of the explicit symmetry breaking,  $V(\phi_0)=0$, without modifying the vacuum
solution, $\phi_0$. In this model $V_0 = -\frac{\tau}{8g^4}$ is chosen to get $V(\phi_0)=0$.
Then the vacuum expectation value of $\Theta^{\nu}_{\nu}$ (which measures the scale symmetry breaking)
vanishes at the condensation, $\phi_0$.
This is the generic structure of the class of Freund-Nambu-type models we will consider below.  One
thing to be noted in this model is that $\phi_0$ is independent of
$\tau$ such that even in the limit of $\tau  \rightarrow 0$ -- that is, when the explicit symmetry
breaking is turned off --  $\phi_0= \frac{1}{g}$ does not change. However if one sets $\tau=0$
{\it ab initio}, the potential is flat and there is no mechanism for the potential to pick the condensate.
This is different from spontaneously broken internal symmetries where the condensate can develop in
the absence of explicit symmetry breaking~\cite{Zumino}. This feature will play an important role in
what follows.

It is interesting to compare the above feature to, say, the
conserved axial current. There chiral symmetry is broken
spontaneously but $\del^\mu A_\mu=0$ with the condensate
$\la\bar{q}q\ra\neq 0$. Here under the scale transformation, for
$\tau\neq 0$, the action is not invariant $\int \delta {\cal
L}_{SB}\neq 0$ but $\del^\mu {D_\mu}|_{\phi_0}=0$, i.e., the
explicit symmetry breaking ``transforms" to spontaneous symmetry
breaking. Note however that the dilaton mass is not equal to
zero.

\subsection{Dilaton model for QCD}
In modelling effective theories of QCD, the dilaton field, $\chi$,
is introduced to take into account of the trace anomaly of QCD
 \be
\theta^{\nu}_{\nu}|_{QCD} = \frac{\beta(g)}{2g}G^a_{\mu\nu}G^{a\mu\nu} +(1+\gamma)m_q\bar{q}q, \label{beta}
 \ee
where $\gamma$ is the anomalous dimension of the scalar $\bar{q}q$, and
$\beta(g)$ is the beta function of QCD.  The basic idea is to introduce the trace anomaly which is an
explicit symmetry breaking as a spontaneously broken scale symmetry in terms of the VEV of the dilaton
as an order parameter on par with the quark condensate as the order parameter for the spontaneously
broken chiral symmetry.

We first consider the chiral limit, setting the quark mass to be zero. In the phase where chiral symmetry is spontaneously broken,  there emerges a mass scale defined by the quark
condensate $\la\bar{q}q\ra$, the order parameter for the spontaneously broken symmetry.  However, the divergence of the dilatation current cannot dictate this scale of quark condensate in the chiral limit, because the second
term vanishes in the chiral limit as in Eq.(\ref{beta}).

What do we know about the
right-hand side of (\ref{beta})?
We know that the condensate $\la G^2\ra\equiv \la G^a_{\mu\nu}G^{a\mu\nu}\ra\neq 0$ in the vacuum
as well as in hot medium up to a temperature $T>T_c$ where $T_c\sim 200$ MeV is the chiral phase
 transition temperature.  Although there are no
model-independent information on this, we expect similar results
in the case of density. In this paper, we assume that it is so. In
fact, in the vacuum $\frac 14
  (\la H|G^2|H\ra-\la 0|G^2|0\ra)$ is the mass of the hadron $H$ (in the chiral limit) -- other
  than the Nambu-Goldstone boson -- which is of course non zero. As will be discussed below,
  $\la 0|G^2|0\ra$ has been measured on the lattice in the vacuum as well as in hot medium and is
  not zero up to a large temperature.

Thus in modelling the trace anomaly of QCD, we expect to have a
different structure from that of the Freund-Nambu model. We will
come back to this problem below. For the moment, let us continue
in the same framework as above.

We write the Lagrangian for the dilaton modelling the trace
anomaly of QCD~\cite{shapo,EL,CEO} in the form \be \frac{1}{2}
\partial_{\mu}\chi
\partial^{\mu}\chi -V(\chi),\label{pot} \ee with \be V(\chi)= B[C + \chi^4
\ln \chi/ \Lambda],\label{CP} \ee where $B$, $C$ and $\Lambda$ are constants
to be determined. The scale dimension of this logarithmic
potential is not four. Both terms break the scale symmetry
explicitly giving the non-vanishing trace of energy-momentum
tensor of the form
 \be \Theta^{\nu}_{\nu} = B[4 C -\chi^4].
 \ee
Now minimizing the potential (\ref{pot}) with respect to $\chi$, one finds the condensate
\be \chi_0 = \Lambda ~
e^{-1/4}. \ee
Suppose one chooses the constant $C$ such that the potential vanishes, $V(\chi_0)=0$. Then one gets
\be
C=\frac 14 \chi_0^4
\ee
which gives the potential of the form
 \be
V(\chi) = B\left[\frac{1}{4} \chi_0^4 + \chi^4
\ln \frac{\chi}{\chi_0 e^{1/4}}\right], \label{ceo}
 \ee
and the trace of the energy-momentum tensor
 \be \Theta^{\nu}_{\nu} = B[\chi_0^4
-\chi^4]. \label{C}
 \ee
This means that the VEV of $\Theta^{\nu}_{\nu}$ vanishes at the condensation $\la\chi\ra=\chi_0\neq 0$:
 \be
\la\Theta^{\nu}_{\nu}\ra|_{\chi_0} =0.
 \ee
This is analogous to the result of the Freund-Nambu model: The explicit symmetry breaking is turned
into a spontaneous breaking of the scale symmetry with the condensate independent of the constant $B$.
This will be what happens with the ``soft" component of the trace anomaly discussed below.
But there is the ``hard" component which remains non-zero to the highest temperature measured so far.
%In what follows below, we will consider the case $C=0$ which is closer to to what we are considering.
%
\section{Chiral Symmetry and Scale Symmetry}\label{CS}
\subsection{``Soft" and ``hard" components}\label{soft-hard}
Although there is no model-independent information on the right-hand-side (RHS) of the energy momentum tensor in dense medium, there is lattice QCD information in hot medium. We assume that we can take over the temperature result to the density case.

What has been measured on the lattice is the gluon
condensate $\la G^2\ra\equiv \la G_{\mu\nu}G^{\mu\nu}\ra$ as a
function of temperature across the chiral transition temperature
$T_c$. Indeed the lattice calculation using dynamical quarks by
Miller~\cite{miller} showed that the gluon condensate starts
melting steeply at what is called ``flash temperature" at about
$T_f\sim 120$ MeV, and loses about a half of the $T=0$ strength at
the chiral phase transition temperature $T_c\sim 200$ MeV above
which the remaining condensate stays flat for some temperature range.
This suggests~\cite{BHHRS,BR-PR} that the gluon condensate
contains, roughly speaking, two components, \be \la G^2\ra_T=\la
G^2_{s}\ra_T +\la G^2_{h}\ra_T \ee with the ``s(oft)" component
vanishing at $T_c$. The lattice result suggests that~\cite{BR-PR}
 \be
\la G^2_{s}\ra_0\approx \la G^2_{h}\ra_0\approx \frac 12 \la G^2\ra_0\label{lattice}
 \ee
and
 \be
\la G^2\ra_{T_c}\approx \la G^2_{h}\ra_{T_c}\approx \frac 12 \la G^2\ra_0.\label{lattice1}
 \ee
There is no fundamental reason why Eqs.~(\ref{lattice}) and (\ref{lattice1}) should hold with equality but in what follows, we will simply take the equal sign for simplicity. The qualitative structure does not depend crucially on it.

These relations suggest to introduce two dilatons $\chi_{s,h}$ for the
soft component and hard component, respectively. If we assume that
there is little mixing between the two dilaton fields as suggested by
Miranksy and Gusynin~\cite{miransky}, then we can write \be
\la\theta^{\nu}_{\nu}\ra=-B(\la\chi_{s}^4\ra
+\la\chi_{h}^4\ra).\label{traceinchi} \ee From (\ref{lattice}), we
infer that \be f_{\chi_s}\approx f_{\chi_h} \ee where
$f_\chi\equiv \la 0|\chi|0\ra$, i.e., the vacuum condensate. This
implies that at the critical point labeled with the subscript ``c"
\be \la\chi_{s}\ra|_{c}=0, \ \ \la\chi_{h}\ra|_{c}\neq
0.\label{chicondensate} \ee It is suggestive to identify
$\chi_{s}$ as the ``quarkish" (Q for short) scalar and $\chi_{h}$
the ``gluonic" (G for short) scalar~\cite{miransky}. Interpreted
in this way, the G term will dominate at large $N_c$ while the Q
term will be suppressed by $1/N_c$. The lattice result described
above suggests that the soft part $\la\chi_s\ra$ ``melts" at the
critical point whereas $\la\chi_h\ra$ remaining more or less
constant across the critical point.

The key idea underlying the notion of Brown-Rho scaling~\cite{BR91} and subsequent developments was
that the manifestation of chiral symmetry in matter below the critical point (i.e., $T_c$, $n_c$)
involved predominantly the Q
component $\chi_{s}$. Now it is still a wide-open question what the Q component is in QCD.
This is because in nature there are five iso-scalar scalar mesons between the mass range $600-1710$ MeV
 that can be mixtures of $\bar{q}q$ and $(\bar{q}q)^2$ configurations. For instance,
  a possible scenario~\cite{rischke} could be that $f_0(600)$ is predominantly of a tetra-quark
  component $(\bar{q}q)^2$ and $f_0(1710)$ is mainly of ``quarkonium" component $\bar{q}q$.
  This would imply that at low temperature or density, the scalar $\chi_s$ would be mainly of the
  tetraquark configuration and close to the chiral restoration  point, it will be of the quarkonium
  configuration. There would be a level crossing at some temperature or density. Near the phase
  transition point, the gluonic scalar could mix strongly with the tetra-quark scalar which is pushed
  up to near the gluonic scalar. This feature is consistent with the possibility of decoupling of the
  $\chi_{h}$ from chiral symmetry proposed below. For our discussion, we need not be specific. The only
  feature we will assume is that the Q and G components could be mixed for certain scalar excitations
  observed in nature. We expect that our qualitative feature remains intact by the possible mixing.

So far we have dealt with the chiral limit with a vanishing
Nambu-Goldstone (pion) mass. The presence of the pion mass -- which breaks chiral symmetry explicitly -- entails
explicit breaking of scale symmetry as given by the second term in
the trace of the energy momentum tensor Eq.~(\ref{beta}). Ignoring
the anomalous dimension, the pion mass term in effective chiral
Lagrangians can be written in the form
 \be
 v^3 {\Tr}(U^\dagger M)+ {\rm h.c}\label{mass0}
 \ee
where $v$ is a constant of mass dimension 1, $M$ is the quark mass matrix and
$U=e^{{\rm i} 2\pi/f_\pi}$, $\pi={\mathbf{\tau}}\cdot {\mathbf{\pi}}/2$.
This term has scale dimension 0. This means that under scale
transformation, the mass term (\ref{mass0}) will contribute to the
trace of the energy momentum tensor (4-0) times the mass term,
which is not correct in the sense that the scale dimension is different from 3 of the second term of (\ref{beta}). In order to remedy this defect, it was
proposed in \cite{EL,CEO} that the mass term in the Lagrangian be
multiplied by a scale dimension 3 object, which using the dilaton field, would be of the form
 \be {\cal
L}_{\chi SB}=(\chi/f_\chi)^3 v^3 {\Tr}(U^\dagger M)+ {\rm h.c}\label{mass1} \ee
with $f_\chi\equiv \la 0|\chi|0\ra$. This has
the right structure in the free-space vacuum giving (\ref{mass0})
when the scalar field is expanded as $\chi=f_\chi+\chi^\prime$.
There will be couplings to the fluctuating scalar $\chi^\prime$
which are subject to experimental tests~\cite{EL}.

According to our strategy, the chiral invariant part of the effective Lagrangian must involve $\chi_s$
at the scale low compared with the chiral scale $\sim 4\pi f_\pi$ and this leads to
the BR scaling for light-quark hadrons -- except for the pion -- predicted in 1991 and
supported by hidden local symmetry in the vector manifestation (HLS/VM for short)~\cite{HY:PR}.
Thus we should take in $\chi$ in (\ref{mass1}) only the soft component $\chi_s$ in hot/dense medium.

\subsection{The Wess-Zumio term}
For flavor number $N_f < 3$, there is no Wess-Zumino term. However in the presence of vector mesons
as in hidden local symmetry formulation~\footnote{Hidden local symmetric vector mesons are expected
to play a crucial role in describing dense hadronic matter. The reason is that baryons appear as
solitons in effective meson field theory and when vector mesons are present together with the pion
field, baryons are coherent states of both the pion and the vector mesons appearing as solitons, i.e.,
skyrmions or instantons. Furthermore the recent development of gravity-gauge duality in hadron structure
suggests that the whole infinite tower of the vector mesons figure crucially in giving the ground
state around which mesons are excited and hence the property of meson excitation in medium will be
strongly effected by the solitonic background as emphasized in ~\cite{BHHRS}.}, there are in general
 four terms in the anomalous parity sector that satisfy homogenous anomaly equation. These are not
 topological but play a crucial role in the structure of baryons as was shown in \cite{PRV-dilaton}.
 We call it hWZ term standing for homogeneous WZ term. It turns out that if one requires vector
 dominance in photon-induced processes involving the hWZ terms and use the equation of motion for a heavy
 $\rho$ field, then the four hWZ terms can be reduced to one term~\cite{meissner}
\be
{\cal L}_{hWZ}=
+\textstyle\frac{3}{2} g \omega_\mu B^\mu\label{wzterm}
\ee
with the baryon current
\be
B^\mu =  \frac{1}{24\pi^2} \varepsilon^{\mu\nu\alpha\beta}
\mbox{Tr}(U^\dagger\partial_\nu U U^\dagger\partial_\alpha U
U^\dagger\partial_\beta U).
\ee
In the standard scaling rule used so far, this term is scale-invariant. However in dense matter,
this term brings havoc to the system as found in \cite{PRV-vector} and will be elaborated below, requiring that as density increases, the pion decay constant $f_\pi$~\footnote{In matter, due to the breaking of Lorentz invariance, the time and space components of the pion decay constant are unequal. However the difference between them appears
at higher orders in the large $N_c$ expansion and so will be ignored here.} must increase instead of
decrease and the in-medium $\omega$ meson mass must increase, both of which being at odds with what's
expected in QCD in general and in the property of nuclear matter in particular. A remedy to this
disease~\cite{PRV-dilaton} is to have this term break scale invariance {\em explicitly in medium} as
 \be {\cal L}_{hWZ}= +\textstyle\frac{3}{2} g
\left(\frac{\chi_s}{f_{\chi_s}}\right)^3 \omega_\mu B^\mu.
\label{wztermp} \ee It has been shown in \cite{PRV-dilaton} -- and briefly recalled below -- that with
this modification, both the pion decay constant and the
$\omega$-meson mass are found to drop as desired as density
increases. Note also that in free space, the Lagrangian predicts
an $\omega\chi_s^\prime\pi^3$ coupling with $\chi_s^\prime$
identified with the lowest-lying isoscalar scalar $f_0(600)$ which
could be measured. However an $\omega$ at rest is kinematically
forbidden to decay into the scalar plus 3 pions, so to test
(\ref{wztermp}), one would have to resort to the process where  $\omega$ is off mass-shell.

\begin{figure}[ht]
 \centerline{\epsfig{file=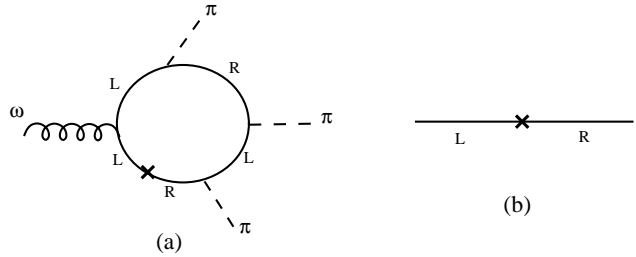,width=14cm,angle=0}}
\vskip -4cm \caption{\small (a) The graph that gives the hWZ
term (\ref{wzterm}) and (b) the graph that give the pion mass term (\ref{mass1}).
The solid line is the L(eft) or R(ight) quark propagator and
the cross stands for the quark condensate $\la\bar{q}q\ra$. Note that the two graphs
share a common feature that carries one factor of the condensate $\la\bar{q}q\ra$ although
(a) should be scale-invariant while (b) breaks scale invariance.}
 \label{anomalygraph}
\end{figure}

There are three ways of justifying the Lagrangian (\ref{wztermp}) as an effective Lagrangian applicable in medium:
\begin{enumerate}
\item One hint is from the behavior of the gauge coupling $g$ in HLS/VM. Near the vector manifestation fixed point, the gauge coupling $g$ goes to zero proportional to $\la\bar{q}q\ra$ as the condensate is dialed to zero at the chiral restoration. The condensate has scale dimension 3, so in terms of $\chi$ field, it transforms as $\chi^3$. Thus we expect in medium
    \be
g^*\approx g(f_{\chi_s}^*/f_{\chi_s})^3\sim g(\la\bar{q}q\ra^*/\la\bar{q}q\ra)
 \ee
\item One can obtain (\ref{wztermp}) from the baryon current derived by Goldstone and
Wilczek~\cite{goldstone-wilczek} appropriate for the Feynman diagram Fig.~\ref{anomalygraph}
(a) which uncovers a ``hidden" quark condensate $\la\bar{q}q\ra$ ``buried" in a
local operator as in the pion mass Fig.~\ref{anomalygraph}(b). Using the sigma model with
the interaction Lagrangian ${\cal L}_I=h\bar{q}(\phi_0+i\vec{\phi}\cdot\vec{\tau}\gamma_5)q$,
Goldstone and Wilczek derived the baryon current of the form
\be
B^\mu\sim \frac{1}{|\phi^4|}\epsilon^{\mu\alpha\beta\gamma}\epsilon_{dabc}\phi_d\del_\alpha
\phi_a\del_\beta\phi_b\del_\gamma\phi_c.
\ee
In the Nambu-Goldstone phase, one can set $\la\phi_0\ra=|\phi|$, so one has
\be
B^\mu \sim \frac{1}{|\phi^3|}\epsilon^{\mu\alpha\beta\gamma}\epsilon_{abc}\del_\alpha
\phi_a\del_\beta\phi_b\del_\gamma\phi_c
\ee
Now this current has scale dimension 6. Therefore coupled to the $\omega$ field, one gets a dimension 7 term corresponding to (\ref{wztermp}).
\item Given that we have two dilaton fields $\chi_s$ and $\chi_h$, one can always form a scale-dimension 0 object $(\chi_s/\chi_h)^n$ for any power $n$. Let us pick $n=3$ and multiply this factor to the scale-dimension-4 term (\ref{wzterm}). In the spontaneously broken phase, we have $\chi_h=f_{\chi_s}+\chi_h^\prime$ where we set $f_{\chi_s}\approx f_{\chi_h}$ as discussed above. Now dropping the ``hard" fluctuation $\chi_h^\prime$, we recover (\ref{wztermp}). What this implies is that there is nothing inconsistent with (\ref{wztermp}) vis-\`a-vis with the scale symmetry property of the WZ term that ``breaks" scale symmetry in medium. Of course multiplying by this factor is totally arbitrary so why it should be cubic instead of any other power is not dictated by any principle. Here, however, we are guided by the above two considerations and perhaps more interestingly by the phenomenological consequence in dense medium found in \cite{PRV-dilaton}. Numerical exercise shows that qualitatively the same result is obtained with $n=2$~\footnote{Private communication from Byung-Yoon Park.}.

\end{enumerate}

\subsection{The model}
To complete the discussion, there then remain terms that are both chirally and scale invariant.
In the absence of the vector fields, they were written down in \cite{EL,CEO}. For instance, the
current algebra term in the chiral Lagrangian takes the form
\be
\frac{f_\pi^2}{4}(\frac{\chi}{f_\chi})^2 {\Tr}(\del_\mu U^\dagger \del^\mu U).
\ee
One can write down other terms of the same class using the standard scale dimension for the vector mesons.

In the spirit of what was discussed above, we need to express our model Lagrangian in terms
of the dilaton $\chi_s$ field only. This can be obtained as follows.

The hidden local symmetry Lagrangian with the appropriate scale transformation can be written (in unitary gauge) as
\be
{\cal L}={\cal L}_{\chi_s}+ {\cal L}_{\chi_h}+{\cal L}_{hWZ}\label{lagtot}
\ee
where
\be
{\cal L}_{\chi_s} &=& \frac{f_\pi^2}{4}\kappa_s^2
%\left(\frac{\chi}{f_\chi}\right)^2
\mbox{Tr}(\partial_\mu U^\dagger \partial^\mu U) + \kappa_s^3 v^3 \mbox{Tr}M(U+U^\dagger)
\nonumber\\
&&
-\frac{f_\pi^2}{4} a\kappa_s^2
%\left(\frac{\chi}{f_\chi}\right)^2
 \mbox{Tr}[\ell_\mu + r_\mu + i(g/2)
( \vec{\tau}\cdot\vec{\rho}_\mu + \omega_\mu)]^2\nonumber\\
&& -\textstyle \frac{1}{4} \displaystyle
\vec{\rho}_{\mu\nu} \cdot \vec{\rho}^{\mu\nu}
-\textstyle \frac{1}{4}  \omega_{\mu\nu} \omega^{\mu\nu}
%\nonumber\\
%&&
%+\textstyle\frac{1}{2} \partial_\mu \chi \partial^\mu \chi
%-\displaystyle \frac{m_\chi^2 f_\chi^2}{4} \left[ (\chi/f_\chi)^4
%(\mbox{ln}(\chi/f_\chi)-\textstyle\frac14) + \frac14 \right]\nonumber\\
+\frac 12 \del_\mu\chi_s\del^\mu\chi_s + V(\chi_s)\label{lags}\\
{\cal L}_{\chi_h}&=&\frac 12\del_\mu\chi_h\del^\mu\chi_h +V(\chi_h)\label{lagh}\\
{\cal L}_{hWZ} &=& \textstyle\frac{3}{2} g \left(\frac{\chi_s}{\chi_h}\right)^3 \omega_\mu B^\mu\label{fwzterm}
\ee
where $\kappa_{s}=\chi_{s}/f_{\chi_{s}}$ and
\be
V(\chi_{s,h})=B\chi_{s,h}^4{\rm ln}\frac{\chi_{s,h}}{f_{\chi_{s,h}}e^{1/4}}.\label{potterm}
\ee
Now letting $\chi_h=f_{\chi_h}+\chi_h^\prime$ and integrating out the $\chi_h^\prime$ field and dropping an irrelevant constant, we have~\footnote{There is a constant term $-\frac{B}{4}f_{\chi_h}^4$ coming from the ``hard" component which we are allowed to drop from the effective Lagrangian we are concerned with since it enters nowhere in the dynamics of the dilaton $\chi_s$. This corresponds to putting $C=\frac{1}{4}f_{\chi_h}^4$ in Eq.~(\ref{CP}) for the $\chi_h$ sector.}
\be
{\cal L}={\cal L}_{\chi_s}+\textstyle\frac{3}{2} g \kappa_s^3\omega_\mu B^\mu.\label{lag2}
\ee
This is exactly the Lagrangian used in \cite{PRV-dilaton}. Note the trace of $\Theta^\mu_\nu$ is of the form
\be
\Theta^\nu_\nu=-B\chi_s^4+\kappa_s^3 v^3{\rm Tr}M(U+U^\dagger).\label{tracesoft}
\ee
As explained above, the {\em apparent} breaking of scale invariance in (\ref{lag2}) does not spoil the scaling property of the model.

We should point out that the second term of Eq.~(\ref{tracesoft}) (or equivalently the second term of (\ref{lags})) {\em dictates} how the scale symmetry breaking given by the condensate $\la\chi_s\ra$ is locked to the chiral symmetry breaking given by the chiral condensate $\la{\rm Tr}(U+U^\dagger)\ra$. In view of what one learns from the Freund-Nambu model, it is clear that this term should play a key role in giving the relation between the two. In what has been studied so far~\cite{BR91,PRV-dilaton}, this term has been treated as a perturbation in the limit the quark mass is taken to be near zero. This may not be reliable for certain quantities, e.g., for the pion. In fact, the Lagranigian (\ref{lag2}) predicts at the mean-field order the pion mass falling as $(\la\chi_s\ra^*/f_{\chi_s})^{1/2}$ in medium but this seems to be at variance with what is seen in experiment~\cite{yamazaki}: The pion mass does not seem to change much either in temperature or density up to the critical point. One could think of this as an indication that the pion is {\em nearly} a Nambu-Goldstone boson, so its mass is protected by chiral symmetry.

It is worth pointing out that the $-\la B\chi_s^4\ra$ term in (\ref{tracesoft}) can be identified with the bag constant of the bag model involved in the chiral phase transition when treated in the mean field of an effective chiral Lagrangian~\cite{BBR}. As indicated in the lattice result~\cite{miller}, it was found to be a half of the total gluon condensate~\cite{BR-PR,BBR}.

It is important to note that $\la\Theta^\nu_\nu\ra$, Eq.~(\ref{tracesoft}), goes to zero in the chiral limit at the chiral symmetry restoration. One could think of this as a Freund-Nambu class, that is, that although away from the chiral transition point, the trace is not equal to zero as we know from lattice, at the chiral transition,
together with quantum (or thermal) contribution terms,  the soft part goes to zero, so that that part of
the trace anomaly can be considered as {\em restored}.  However, the hard part, i.e., the
glue part that in the approximation taken in the paper is unconnected to chiral symmetry, which has to do with the fact that scale symmetry is not really restored in QCD, remains non-zero.
\subsubsection{Implications on mass scaling}
Treated at the tree order or in the mean field as was done in 1991~\cite{BR91}, the Lagrangian would give the scaling of the $\rho$ meson
\be
\Phi(n)\equiv \frac{m_\rho^*}{m_\rho}\approx \frac{f_{\chi_s}^*}{f_{\chi_s}}\approx
\frac{f_{\pi}^*}{f_{\pi}}.
\ee
This is at variance with the HLS prediction near the vector manifestation fixed point~\cite{HY:PR} which predicts
\be
\Phi(n) \approx \frac{\la\bar{q}q\ra^*}{\la\bar{q}q\ra}\approx \left(\frac{f_{\chi_s}^*}{f_{\chi_s}}\right)^3.
\ee
This discrepancy is not surprising. The HLS approach is a renormalization-group approach -- though in practice at  one-loop order -- on the one hand and involves an uncontrolled method in treating density effects on the other. In HLS in the vector manifestation, it is the gauge coupling $g$, not the pion decay constant $f_\pi$, which scales as $\sim \la\bar{q}q\ra$. This means that the identification of the HLS prediction in terms of the mean-field contribution from Eq.~(\ref{lag2}) done here cannot be reliable near the chiral transition. As will be mentioned below, when put on crystal lattice to simulate density, there is a qualitative change in the structure of dense matter. Phenomenological considerations in nuclear structure indicate that the scaling behavior changes both in density and temperature from below the ``flash point" (density $n_f > n_0$ and temperature $T_f\sim 120$ MeV) to above, with a rough dependence subject to a ``double decimation"~\cite{DD}, $\Phi$ going from $\sqrt{\la\bar{q}q\ra^*/\la\bar{q}q\ra}$ to $(\la\bar{q}q\ra^*/\la\bar{q}q\ra)$. Given the complexity and multiple scales involved in nuclear processes, it would be oxymoronic to take either one or the other blindly and apply to hadronic matter in the normal as well as extreme conditions without carefully checking whether the approximations make sense in the regime involved.

Despite all these caveats, it is an interesting possibility that the local gauge fields in HLS  can be considered as the $\psi$-type ``matter field" in the Freund-Nambu model. The HLS fields (by construction) enter scale-invariantly but their masses go to zero as the scale invariance of the soft sector is restored, that is, as $\la\chi_s\ra\rightarrow 0$. This is analogous to that both the mass of the Freund-Nambu dilaton $\phi$ and that of the ``matter field" $\psi$ go to zero as the condensate $\phi_0$ is dialed to zero.
\section{Holographic Dense Matter}
\subsection{The problem}
Cold dense baryonic matter relevant to the physics of the interior of compact stars is still poorly understood. At superhigh density, the asymptotic freedom allows QCD to predict that color superconductivity in the form of color-flavor locking will form the ground state of the system. However at densities pertinent for compact stars where the interaction coupling is not weak, no reliable model-independent theoretical tools applicable under realistic conditions are available.

Given this situation, aside from the intrinsic interest on its own merit, it is natural that a great deal of attention is currently paid to the gravity-dual approach to QCD -- referred to as holographic dual QCD -- that is thought to have the potential to address the strong coupling regime involved. The phenomenon we are specifically interested in for nuclear matter as well as compact-star matter must involve both chiral symmetry and confinement-deconfinement. The appropriate model which implements both these features correctly in string theory has recently been constructed by Sakai and Sugimoto (SS for short)~\cite{sakai} using D4/D8 branes. Since this is currently the only model that has correct chiral symmetry of QCD, we will focus on this model. There is a large body of literature on hot or cold dense matter in terms of various different holographic models, but most of them do not address the problem we are concerned with here. We will not discuss them here although they are quite illuminating for a variety of other phenomena such as in heavy ion collisions.

There have been detailed studies of dense nuclear matter in the SS model~\cite{kimetal,bergmanetal,rozalietal,holo-nuclei,fermisurface}. What interests us here is the set of predictions made in this model, in particular on the properties of hadrons in medium, in the range of densities relevant to compact stars, i.e., from that of the nuclear matter to the chiral transition density. This is the range that will be probed experimentally in such accelerators as FAIR/GSI.

Now the SS model is valid for the limits $N_c\rightarrow \infty$, $\lambda\equiv g^2 N_c\rightarrow \infty$ (i.e., 't Hooft limit), $N_f/N_c\ll 1$ (``probe approximation") and in the chiral limit. This is roughly the regime in which the quenched approximation in lattice QCD applies and indeed many of the hadron properties that are given reliably in (quenched) lattice QCD calculations can be reproduced in the SS model~\cite{sakai,HRYY, hataetal,hataetal2,hashimotoetal,Kim-Zahed}. In contrast to this successful confrontation with Nature in matter-free space, the result obtained in \cite{kimetal,bergmanetal,rozalietal} indicates however that there is something basically missing in the SS model when applied to nuclear and denser matter. For instance, the prediction is that both space and time components of the pion decay constant $f_\pi^{s,t}$ {\em increase} as density is increased which is at variance with what is expected in QCD where the pion decay constants should decrease, going to zero at $n_\chi$. This is also at odds with Brown-Rho scaling~\cite{BR91}.

The remedy brought by the model derived above in the HLS case, Eq.~(\ref{lag2}), helps identify the source of this behavior in the holographic description.
\subsection{Dense matter with the HLS model}
To bring out the basic problem, we first discuss dense matter in HLS theory without dilatons. Here we will encounter the same problem as the one met in the holographic model.

In the unitary gauge, we write the HLS Lagrangian as
\begin{eqnarray}
{\cal L} &=& \frac{f_\pi^2}{4}
%\left(\frac{\chi}{f_\chi}\right)^2
\mbox{Tr}(\partial_\mu U^\dagger \partial^\mu U) + v^3
%\left(\frac{\chi}{f_\chi}\right)^3
    \mbox{Tr}M(U+U^\dagger)
\nonumber\\
&&
-\frac{f_\pi^2}{4} a
%\left(\frac{\chi}{f_\chi}\right)^2
 \mbox{Tr}[\ell_\mu + r_\mu + i(g/2)
( \vec{\tau}\cdot\vec{\rho}_\mu + \omega_\mu)]^2\nonumber\\
&& -\textstyle \frac{1}{4} \displaystyle
\vec{\rho}_{\mu\nu} \cdot \vec{\rho}^{\mu\nu}
-\textstyle \frac{1}{4}  \omega_{\mu\nu} \omega^{\mu\nu}
%\nonumber\\
%&&
%+\textstyle\frac{1}{2} \partial_\mu \chi \partial^\mu \chi
%-\displaystyle \frac{m_\chi^2 f_\chi^2}{4} \left[ (\chi/f_\chi)^4
%(\mbox{ln}(\chi/f_\chi)-\textstyle\frac14) + \frac14 \right]\nonumber\\
+ {\cal L}_{hWZ},
\label{lag-nochi}\end{eqnarray}
with
\be
{\cal L}_{hWZ}=
+\textstyle\frac{3}{2} g \omega_\mu B^\mu.\label{wzterm-nochi}
\ee
As mentioned, this hWZ term follows from making certain approximations -- such as vector dominance in the photon-induced process and massiveness of the $\rho$ meson~\cite{meissner}. It appears in an on-going work~\cite{oh} that the qualitative result remains the same when such approximations are not made. What seems to matter is that the hedgehog $\rho$ meson carries baryon density and hence can replace the pionic baryon charge density $B_0$. It thus seems reasonable to consider (\ref{wzterm-nochi}) as characteristic of the anomaly term.

Skyrmions in the model (\ref{lag-nochi}) have been put on an FCC crystal lattice to simulate dense matter~\cite{PRV-vector}. The result has been found to be quite similar to what was found in the holographic model. It is easy to see what happens.

The essential point is that the $\omega$ meson gives rise to a Coulomb potential. The hWZ term then leads to the repulsive interaction, contributing to the energy per baryon, $E/B$, of the form
\be
(E/B)_{WZ}
 = \frac{9g^2}{16}\textstyle \displaystyle
\int_{\mbox{\scriptsize Box}} d^3x \int d^3 x^\prime
B_0(\vec{x}) \frac{e^{-m_\omega^*}
|\vec{x}-\vec{x}^\prime|} {4\pi|\vec{x}-\vec{x}^\prime|}
B_0(\vec{x}^\prime). \nonumber
\ee
What is important is that this repulsive interaction turns out to dominate over other terms as density increases. Now while the integral over $\vec{x}$ is defined in a single lattice (FCC) cell, that over $\vec{x}^\prime$ is not, so will lead to a divergence unless tamed. In order to prevent the $(E/B)_{WZ}$ from diverging, $m_\omega^*$ has to {\em increase} sufficiently fast. And since $m_\omega^*\sim f_\pi^* g$ in this model, for a fixed $g$, $f_\pi^*$ must therefore {\em increase}. This phenomenon is a generic feature associated with the role that the vector mesons in the $\omega$ channel play in dense medium. This feature, however, is at variance with nature: QCD predicts that the pion decay constant tied to the chiral condensate should {\em decrease} and go to zero (in the chiral limit) at the chiral transition.

The remedy to this disease turns out to be provided by the dilaton-implemented model Eq.~(\ref{lag2}) with, most importantly, the modified hWZ term. In fact if one drops the $\kappa_s^3$ factor in the hWZ term while keeping the $\kappa_s$ dependence in the normal component of the Lagragian, one finds little qualitative difference from the result of the Lagrangian without coupling to the dilatons, (\ref{lag-nochi}).

The effect of dilatons in the form of Eq.~(\ref{lag2}) turns out to be dramatic for both single skyrmion as well as dense skyrmion matter~\cite{PRV-dilaton}. There is a qualitative difference in physics between ``light" dilaton (LD) $m_{\chi} < 4\pi f_\pi\sim 1$ GeV and ``heavy" dilaton (HD) $m_\chi \gsim 1$ GeV.~\footnote{In what follows, unless otherwise stated, by dilaton, we will understand the soft component $\chi_s$ only.} We do not have a clear understanding of what causes the drastic difference nor of the precise mass that delineates the two regimes. As noted in Section \ref{soft-hard}, there is a likely level-crossing between the low-density and high-density regimes of the quarkonic (i.e., $q\bar{q}$) and tetra-quark $(q\bar{q})^2$ configurations which would make the identification of the $\chi$ field subtle. It is not clear that the non-mixing assumption between the soft and hard dilatons that is made in the discussion could be reliable. These issues are to be considered as caveats to the results.

What came out may however be qualitatively robust.

Roughly speaking, a single skyrmion with a light dilaton (LD) resembles the ``Little Bag"~\cite{littlebag} with a small ``confinement size" while a heavy dilaton (HD) resembles that of an MIT bag with a big confinement size. Whether and how this drastic difference can be ``gauged away" as implied in the Cheshire Cat principle remains to be studied.

\begin{figure}[hbtp]
\centerline{\epsfig{file=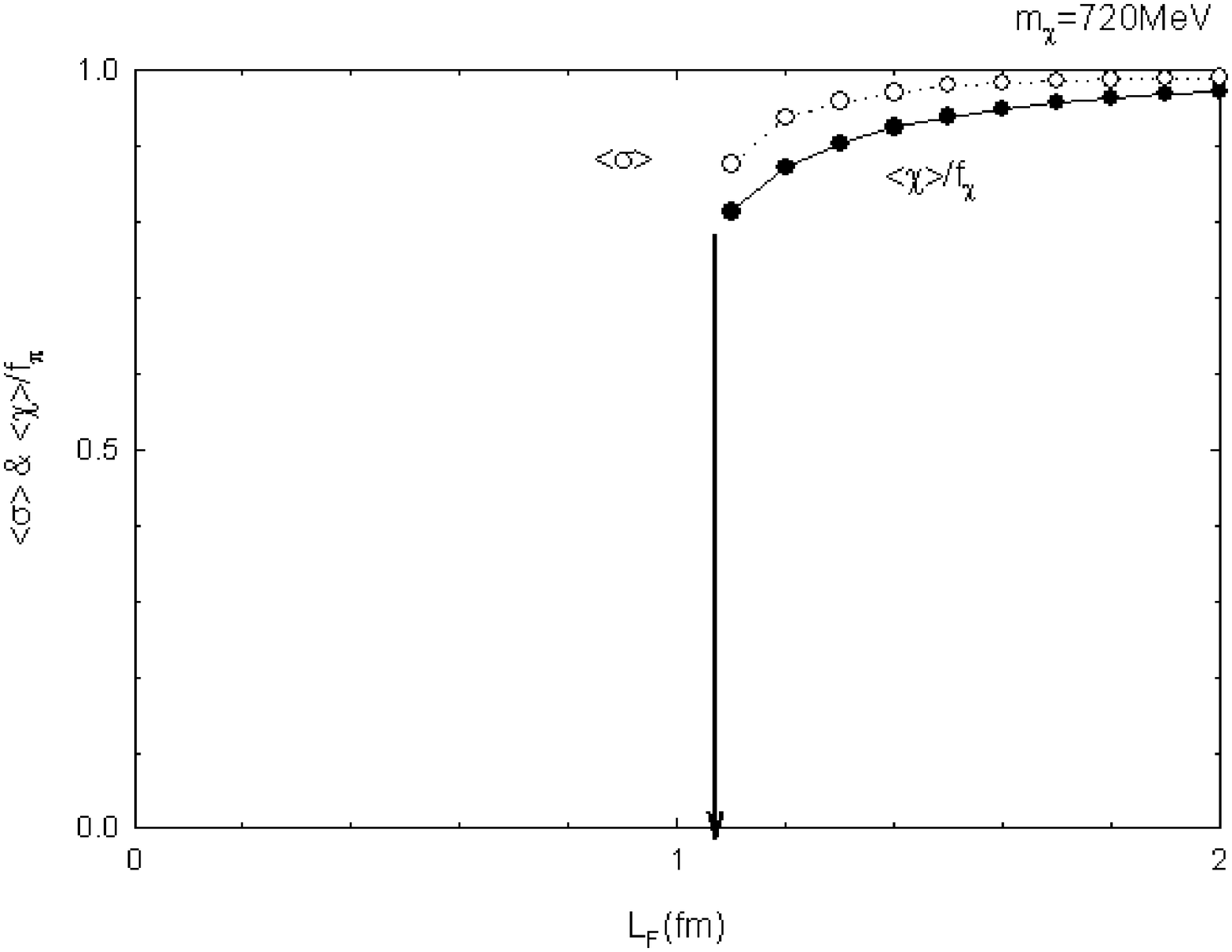,width=7cm,angle=0}
\epsfig{file=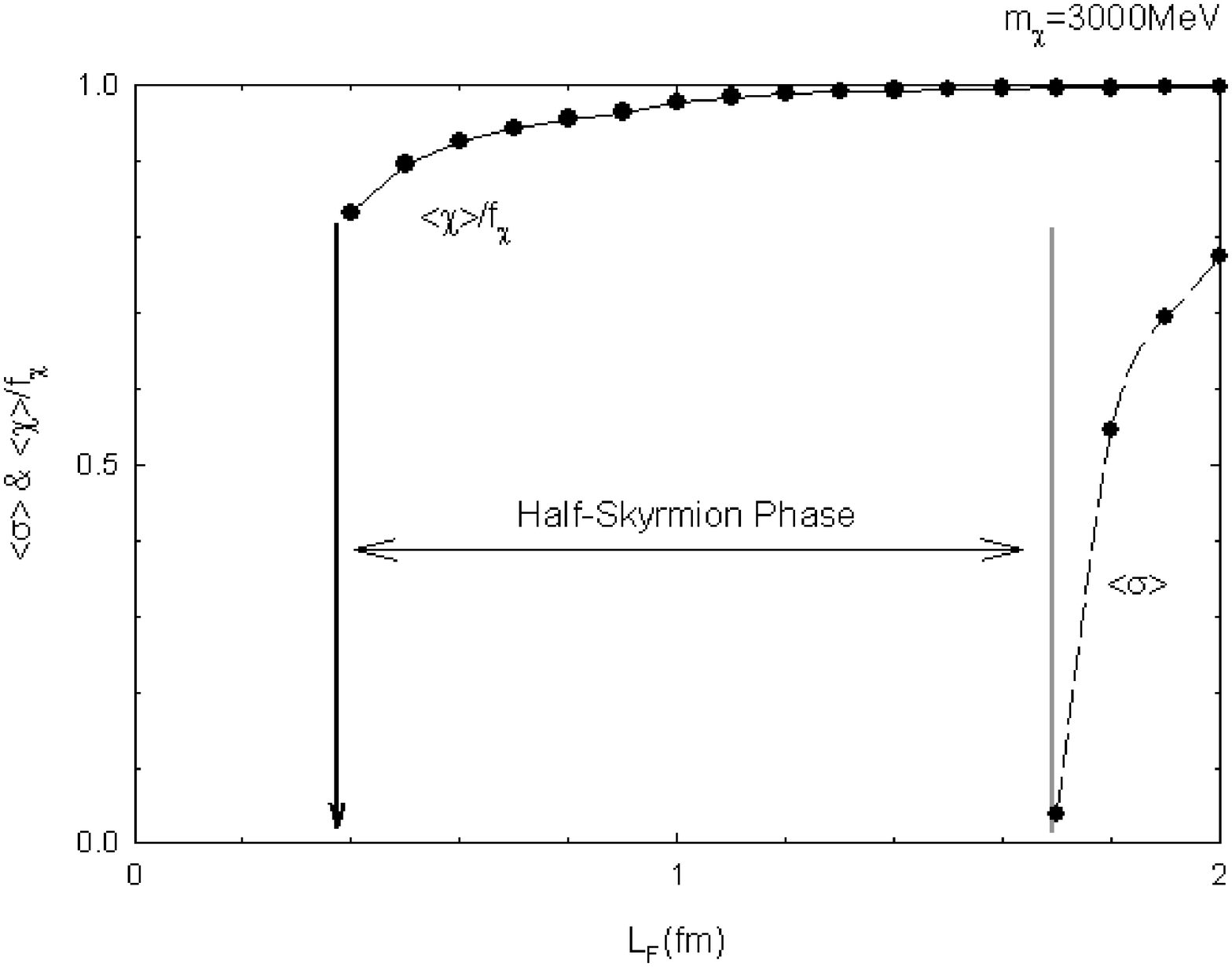,width=7cm,angle=0}}
\caption{Behavior of
$\langle\chi\rangle$ and $\langle\sigma\rangle\propto \la\bar{q}q\ra$ where $\sigma=\frac 12 {\rm Tr}U$ as a function
of lattice size for ``light" dilaton mass $m_\chi=720$ MeV (left figure) and for ``heavy" dilaton mass
$m_\chi=3000$ MeV (right figure).}
\label{half-skyrmion}
\end{figure}

A more remarkable difference -- which is highly relevant to this part of the discussion -- takes place in the phase structure in density. The result of the calculation in \cite{PRV-dilaton} is shown in Fig.~\ref{half-skyrmion}. With a HD, the skyrmion matter makes a phase transition at a relatively low density to a half-skyrmion matter with a vanishing quark condensate $\la\bar{q}q\ra^*\rightarrow 0$ but with $f_\pi^*\sim \la\chi\ra^* \neq 0$. This phase corresponds, in HLS theory, to the phase with $a=1$ (i.e., $f_\pi=f_\sigma\neq 0$ in the notation of \cite{HY:PR}) and $g\neq 0$ but $\la\bar{q}q\ra=0$. This means that in this phase, chiral symmetry is restored but quarks are not deconfined. Since this transition takes place in a theory valid at large $N_c$, it can be identified with what McLerran and Pisarski call ``quarkyonic phase"~\cite{McLerran}. We suggest that this half-skyrmion phase, naturally realized in HLS skyrmions, is generic at high density~\cite{MR:half}.

Now when the mass of the dilaton is light, the half-skyrmion phase shrinks to a point at which HLS ceases to be valid. This can be interpreted as the half-skyrmion phase wedged between the confinement and deconfinement regimes as well as between the Nambu-Goldstone and Wigner-Weyl phases.

Within the given approximations and with the caveats mentioned above, it is not clear whether this sharp delineation between the two scenarios is real in the same manner as the contrasted skyrmion structure observed in a single skyrmion could not be dismissed as an artifact. Much work is needed to clarify this issue.

Whatever the case is, what is significant is that with the dilaton-coupled hWZ term, both the pion decay constant and the $\omega$ mass {\em do drop} as a function of density as is expected in QCD.
\subsection{Nuclear effective field theory}
It is instructive to translate the description in terms of skyrmions for nuclear and dense matter discussed above into a nuclear field theory framework. Such a translation has not been explicitly worked out in the context of hidden local symmetry theory. However there is nothing that suggests that it cannot be done. The framework that could lead to such a feat based on Weinberg's ``folk theorem"~\cite{weinberg-theorem} is detailed in the monograph \cite{cndII}.

In the present context, Weinberg theorem implies the following. Instead of working with a realistic HLS Lagrangian given in terms of meson fields only from which baryons arise from skyrmions after suitable collective-coordinate (or moduli) quantization, one posits nucleons as local fields and constructs, in consistency with all relevant symmetries, effective field theories for nuclear physics in terms of nucleons, pions, vector mesons etc. that comprise the relevant degrees of freedom for the given scale and do a systematic perturbation theory built on a Lagrangian that has the appropriate symmetries. In low-energy nuclear physics, this strategy is exemplified by heavy-baryon chiral perturbation theory~\cite{weinberg-chpt}.

Adopting this strategy, one can understand the problem (with $f_\pi$ etc.) discussed above from the point of view of standard effective nuclear field theory in terms of a Lagrangian written with baryon field treated in mean field. Let us consider, for simplicity, symmetric nuclear matter. In this case, the relevant Lagrangian is Walecka's linear mean field Lagrangian~\cite{walecka},
 \be
{\cal L}_{walecka} &=&
\bar{N}(i\gamma_{\mu}(\del^\mu+ig_v^\star\omega^\mu
)-M^\star+h^\star\phi )N \nonumber\\ & &-\frac 14 F_{\mu\nu}^2
+\frac 12 (\partial_\mu \phi)^2
+\frac{{m^\star_\omega}^2}{2}\omega^2
-\frac{{m^\star_\phi}^2}{2}\phi^2+\cdots\label{leff2}
 \ee
Here the ellipsis stands for higher-dimension operators allowed by the symmetries involved. The asterisk stands for in-medium quantities that need be specified.  The isovector pion and vector fields are not written down since they do not contribute in the mean field to symmetric nuclear matter. Here the scalar $\phi$ -- which is a chiral scalar -- plays the same role as the dilaton scalar $\chi_s$.~\footnote{There have been discussions on a possible relation between the dilaton $\chi_s$ that enters in the spontaneous breaking of scale invariance and the scalar ``$\sigma$" that provides attraction in nuclear matter, the vev of which ultimately represents the chiral order parameter $\la\bar{q}q\ra$ as in the linear sigma model~\cite{beane-vankolck,cndII}.} The $\omega$ field is also a chiral scalar, so the Lagrangian has the correct chiral symmetry as (\ref{lag2}). Treated in the mean field, this theory should be equivalent -- in the sense of Weinberg's folk theorem as stated above -- to that of the skyrmion matter treated with the Lagrangian (\ref{lag2}). Furthermore, the Walecka mean field theory is equivalent to Landau Fermi liquid theory~\cite{matsui} and hence gives a good description of nuclear matter at its saturation point. If one takes all the quantities with asterisk to be constant independent of density and fixed at zero density, this Lagrangian in mean field can describe nuclear matter by fine-tuning the parameters~\footnote{The compression modulus is a different matter but can be suitably corrected with the parameters scaling \`a la Brown-Rho scaling~\cite{song}.}. The nuclear matter saturation occurs via a delicate balance between the vector mean field $\la\omega_0\ra$ and the scalar mean field $\la\phi\ra$, the former repulsive and the latter attractive. As density increases, the vector repulsion going as $\la N^\dagger N\ra$ increases {\em faster} than the scalar attraction going as $\la\bar{N}N\ra$. The consequence is that chiral restoration signalled by the nucleon mass going to zero can take place {\em only when} the density goes to infinity. With the negative energy states taken into account, the mean field theory does not even saturate~\cite{kochetal}.~\footnote{In this reference, Walecka mean field theory is formulated in terms of nucleon fields only, containing higher dimension operators $(\bar{N}N)^n$ with $n\geq 2$. This is equivalent to doing mean field with (\ref{leff2})which is also equivalent to Landau Fermi-liquid fixed point theory~\cite{cndII}.} What is required to mend this defect is higher dimension operators consistent with chiral symmetry. These higher dimension terms can be interpreted as a part, if not the whole content, of the ``intrinsic density dependence" lodged in HLS Lagrangian arising from the Wilsonina-matching to QCD in medium~\cite{HY:PR} and equivalently as BR scaling~\cite{song}.

Describing chiral phase transition in Walckea mean field approach is discussed in \cite{BBR}. What transpires from the analysis of this work in view of what was discussed above for (\ref{lag2}) is that there are two ingredients that are essential: One is the scalar degree of freedom represented by the dilaton field $\chi_s$ associated with the spontaneous breaking of scale invariance figuring in the skyrmion crystal (\ref{lag2}) and the scalar $\phi$ in Walecka model (\ref{leff2}) that provides attraction in nuclear interactions as well as a ``bag constant" in the vacuum; the other is the vector manifestation in HLS theory that drives the vector coupling $g$ toward zero, the key mechanism for Brown-Rho scaling. In the skyrmion description, the latter is reflected in the dilaton modified hWZ term.
\subsection{Dense nuclear matter in the Sakai-Sugimoto model}
The Sakai-Sugimoto action~\cite{sakai} valid in the large $N_c$ and $\lambda$ limit and in the probe (or quenched) approximation can be reduced to the form~\cite{HRYY}
\be
S=S_{YM} +S_{CS}\label{SSaction}
\ee
where
\begin{equation}
S_{YM}=-\;\int dx^4 dw
\;\frac{1}{2e^2(w)} \;{\rm tr} F_{mn}F^{mn}+\cdots\:,\label{YM}
\end{equation}
with $(a,b)=0,1,2,3,w$ is the 5D YM action coming from DBI action
where the contraction is with respect to the flat metric $dx_\mu dx^\mu+dw^2$
and the position-dependent electric coupling  $e(w)$ is given by
\begin{equation}
\frac{1}{e^2(w)}
\simeq \frac{(g_{YM}^2N_c)N_c}{108\pi^3}M_{KK}(1+M_{KK}^2 w^2)
\end{equation}
and
\begin{equation}
S_{CS}=\frac{N_c}{24\pi^2}\int_{4+1}\omega_{5}(A)
\end{equation}
with $d\omega_5(A)=\tr F^3$ being the Chern-Simons action that encodes anomalies. The action (\ref{YM}) involves a series of additional approximations made to the form obtained by SS~\cite{sakai}. The details are irrelevant here. What is relevant for the discussion that follows is its generic 5D form.

With the low-energy strong interaction dynamics given in the form (\ref{SSaction}), baryons arise as instantons from $S_{YM}$~\footnote{When KK-reduced to 4D, the action $S_{YM}$ consists of an infinite tower of hidden local vector and axial vector gauge fields in addition to the pion field, and baryons can then be described in terms of skyrmions in the presence of the infinite tower.} whose size shrinks to zero as $\sim \lambda^{-1/2}$ as $\lambda\rightarrow \infty$~\cite{HRYY,hataetal,hashimotoetal}\footnote{The physical size of the baryon should not depend on $\lambda$~\cite{HRYY,hashimotoetal}, so the instanton size that shrinks as $1/\sqrt{\lambda}$ is not physical. As pointed out by Kim and Zahed~\cite{Kim-Zahed}, this independence of the size on $\lambda$ is just an aspect of what is known as the Cheshire cat phenomenon~\cite{cndII}.}. The shrinking of the instanton is prevented by the Chern-Simons term which takes the form in the presence of the instanton
\be
S_{CS}=\frac{N_c}{8\pi^2}\int A_0{\tr}(F\wedge F).\label{cs}
\ee
The Chern-Simons term provides a Coulomb repulsion that stabilizes the instanton. As noted in \cite{hataetal}, this stabilization is quite analogous to the $\omega$ meson stabilization of the skyrmion \`a la Adkins and Nappi~\cite{adkins-nappi}.

In applying the action (\ref{SSaction}) to many-baryon systems, it would be possible to put multi-instantons on crystal lattice as was done with skyrmons in \cite{PRV-dilaton}. In doing so, one should note that the instanton density distribution ${\tr}(F\wedge F)$ in (\ref{cs}) is nothing but the baryon number density distribution and hence $A_0$ can be interpreted as a variable conjugate to density, i.e., the baryon chemical potential $\mu_B$. This term is a 5D analog to the hWZ action with (\ref{wzterm}), playing the same role as the $\omega$ stabilization in the skyrmion model. Such a dense instanton crystal or equivalently a crystal of skyrmions in the presence of an infinite tower of vector mesons, although not yet constructed, would have a similar property as the HLS crystal discussed above.

The analogy can be made even more apparent by going to an effective action of (\ref{SSaction}). To do this, we note first that the properties of the elementary nucleon that are reliably calculable in the quenched approximation in lattice QCD can be more or less reproduced at the tree level with (\ref{SSaction})~\cite{sakai,HRYY, hataetal,hataetal2,hashimotoetal,Kim-Zahed}. This can be rephrased in terms of an ``effective action" consisting of an isospin 1/2 Dirac field ${\cal B}$ for the five-dimensional baryon~\cite{HRYY}
\begin{eqnarray}
S&=& -\int d^4 x  dw {\frac{1}{2} e^{-2}(w)} \;{\tr}\, F_{mn}F^{mn}\nonumber\\
&+&\int d^4 x dw [-i\bar{\cal B}\gamma^m D_m {\cal B}
-i m_b(w)\bar{\cal B}{\cal B} \nonumber\\
&& +g_5(w){\rho_B^2\over
e^2(w)}\bar{\cal B}\gamma^{mn}F_{mn}{\cal B} ]+\cdots\,,
\label{5dfermion1}
\end{eqnarray}
where $\rho_B$ is the stabilized size of the 5D instanton
representing baryon, and $g_5(w)$ is a function whose
value at $w=0$ can be determined from the instanton structure of the theory, $g_5(0)=2\pi^2/3$. Not only the coefficients of this action but also various static properties of the nucleon are calculated via the soliton structure. This action can be made suitably applicable to many-nucleon systems. This procedure of writing down an effective action in terms of relevant hadronic variables and applying it to dense matter is totally analogous to capturing multi-skyrmion physics by baryon chiral perturbation theory introduced by Weinberg for nuclear physics in terms of the 4D baryon field $B$~\cite{weinberg-chpt,cndII}. Going from hidden local symmetry Lagrangian~\cite{HY:PR} given in terms of the Nambu-Goldstone pion field $\pi$ and the vector fields $\rho$ and $\omega$ (for $N_f=2$) via solitons to Weinberg-type Lagrangian for effective field theory of nuclei and nuclear matter is implicit along the line of Weinberg's ``folk theorem"~\cite{weinberg-theorem}.

We know from Walecka theory that doing a mean-field calculation with the action (\ref{5dfermion1}) should be a lot more realistic than what was done in \cite{kimetal,bergmanetal,rozalietal} with the mesonic action (\ref{SSaction}) which roughly represents a mean-field treatment of the meson fields of the infinite tower with the baryon charge sourced by the Chern-Simons term, with density distributed uniformly. In the latter, the inter-baryon correlations present in the multi-skyrmion configurations -- which are captured largely by the explicit presence of the baryon field in the Walecka mean-field approach -- is absent. The point we are making is that even a Walecka-type treatment with (\ref{5dfermion1}) would suffer from the defect. Our conjecture is that the remedy to the disease in holographic dense matter could be found in the kind of scalar degrees of freedom encountered in HLS theory. This point is strengthened by the observation that HLS with the lowest-lying vector mesons (HLS$_1$) is holographic HLS theory with the infinite tower HLS$_\infty$ in which {\em all} except the lowest vector mesons are integrated out~\cite{HMY}.

\section{Conclusion}
By identifying the two component structure of the QCD trace anomaly with a ``soft" dilaton and a
``hard" dilaton and introducing a scale-invariant Wess-Zumino term that effectively scales with the
soft dilaton, it is found to be possible to make the connection between the scaling of hadron properties
based on Brown-Rho scaling which was originally anchored on spontaneous breaking of scale invariance in
medium and that based on renormalization group flow of hidden local symmetry in the vector manifestation.
The vanishing of the vector meson mass at the vector manifestation fixed point in HLS can be related to
the dilaton condensate going to zero representing a spontaneously broken scale symmetry being
{\em restored}. We should stress that had we used a single dilaton for the trace anomaly as in \cite{EL,CEO}, we would have encountered two major problems. One is that we would not have been able to construct an hWZ term in consistency with the scaling invariance that scaled in density as we needed for decreasing $f_\pi$ and $\omega$ mass. The other is that Brown-Rho scaling would not have matched with the HLS in the vector manifestation. The $\rho$ mass would not have scaled to zero at chiral restoration, simply because $\Theta_\mu^\mu$ does not vanish across the critical point~\footnote{This aspect was noted early on~\cite{birse} when the scaling relation was first proposed in \cite{BR91}.}.

The concept of the ``flash" point that figures in the description of certain heavy-ion proceeses (flash temperature $T_f$) and EOS for neutron-star matter (flash density $n_f$)~\cite{BHHRS} is anchored on the soft dilaton.

It is found that the two component dilaton structure which renders the
scaling behavior of the homogeneous Wess-Zumino term of the
$\omega\cdot B$-type consistent with scale invariance of the Lagrangian plays an extremely
important role in the structure of the nucleon, nuclear and dense
matter as found in \cite{PRV-dilaton}. Depending upon the dilaton mass -- light or heavy, the nucleon can be described as a ``Little Bag" or as a big MIT bag. More significantly, in dense matter, both the pion decay constant and the $\omega$ meson mass are found to decrease as density increases, with a half-skyrmion phase emerging at high density that can be identified as the ``quarkyonic phase" predicted at large $N_c$~\cite{McLerran}.

We should remark that the form of the hWZ term (\ref{wzterm}) is
of a special form in HLS theory obtained when vector dominance is
first imposed on all photon-induced processes and then the $\rho$
field is eliminated by means of its equation of motion in the limit of large mass. In general
there are four terms in the hWZ Lagrangian involving both the
$\rho$ and $\omega$ fields with an arbitrary set of coefficients.
The recent development of holographic QCD~\cite{sakai} suggests a
set of coefficients that would assure vector dominance in the
$\omega\rightarrow 3\pi$ channel which would make the
$\omega\cdot B$-type term drop out. This is highly relevant since in HLS theory where the vector meson mass is to be treated on the same footing as the pion mass, the large mass limit used to arrive at the $\omega\cdot B$-type hWZ term is suspect. The question arises what happens to the
scaling behavior of the more general hWZ terms in medium, in particular with
the behavior of the vector meson, the pion decay constant etc. A
diagrammatic consideration presented above indicates, however, that
the same $\kappa_s^3$ factor will intervene and bring a similar
qualitative feature as with (\ref{wzterm}). This issue needs to be
studied~\cite{oh}.

One might wonder whether there could not be arbitrariness in exploiting the two-component dilatons. While the scale invariant combination $(\chi_s/\chi_h)^n$ with $n=3$ is invoked for the reasons detailed above for the hWZ term, one might imagine multiplying -- albeit without a compelling reason for it -- by similar factors with an arbitrary $n$ the normal component of the HLS Lagrangian without affecting either the hidden local symmetry or the chiral symmetry. That would, however, generate an arbitrary scaling behavior in dense matter. Such an arbitrariness could -- and should -- be eliminated by checking the coupling coefficients of the (fluctuating) $\chi_s^\prime$ field to other (matter) fields as discussed in \cite{EL}. This issue needs to be further investigated.

One of the potential spin-offs of the role of dilatons is a possible remedy to the serious defects found in the holographic approach to dense matter. The mechanism associated with the hWZ
term that makes the $\omega$ mass as well as the pion decay
constant drop in hot/dense medium as expected in QCD could also
resolve the problem faced in the holographic approach to dense
matter where the opposite behavior was found. In the Sakai-Sugimoto model, this requires a proper introduction of scalar degrees of freedom representing spontaneous breaking of scale symmetry which gets restored at the chiral restoration point. How to introduce such degrees of freedom with the D4/D8 brane dynamics is at present unknown.

Finally the way the $\omega$ repulsion is tamed in dense matter as described in this paper can have ramifications on the equation of state for neutron stars and the formation of black holes. For instance, the untamed $\omega$ repulsion that forces the $f_\pi$ and the $\omega$ mass to increase at increasing density would ``banish" kaon condensation to a density above $\sim  7n_0$, making the Brown-Bethe maximum neutron star mass $M^{max}\simeq 1.56 M_\odot$ untenable~\cite{BLR-kaon}. This matter will be discussed elsewhere.

\subsection*{Acknowledgments}
We are grateful for discussions with Gerry Brown, Masa Harada, Byung-Yoon Park and Vicente Vento on various aspects
of the problem discussed in this article. This work is supported by the WCU project of Korean Ministry of Education, Science and Technology (R33-2008-000-10087-0).

{}

\end{document}